\documentclass{article}

\usepackage{microtype}
\usepackage{graphicx}
\usepackage{booktabs} 
\usepackage{floatrow}
\usepackage{subcaption}
\usepackage{amsfonts}
\usepackage{floatrow}
\usepackage{hyperref}

\usepackage[accepted]{icml2018}

\icmltitlerunning{Towards Gene Expression Convolutions using Gene Interaction Graphs}

\begin{document}

\twocolumn[
\icmltitle{Towards Gene Expression Convolutions using Gene Interaction Graphs}

\icmlsetsymbol{equal}{*}

\begin{icmlauthorlist}
\icmlauthor{Francis Dutil}{equal,mila}
\icmlauthor{Joseph Paul Cohen}{equal,mila}
\icmlauthor{Martin Weiss}{mila}
\icmlauthor{Georgy Derevyanko}{cermm}
\icmlauthor{Yoshua Bengio}{mila}
\end{icmlauthorlist}

\icmlaffiliation{mila}{Montreal Institute for Learning Algorithms, Universit\'{e} of Montr\'{e}al}
\icmlaffiliation{cermm}{Centre for Research in Molecular Modeling (CERMM), Concordia University}

\icmlcorrespondingauthor{Francis Dutil}{dutilf@iro.umontreal.ca}
\icmlcorrespondingauthor{Joseph Paul Cohen}{cohenjos@iro.umontreal.ca}

\icmlkeywords{gene expression, RNA-seq, graph convolutions}

\vskip 0.3in
]



\printAffiliationsAndNotice{\icmlEqualContribution} 

\begin{abstract}
We study the challenges of applying deep learning to gene expression data. We find experimentally that there exists non-linear signal in the data, however is it not discovered automatically given the noise and low numbers of samples used in most research.
We discuss how gene interaction graphs (same pathway, protein-protein, co-expression, or research paper text association) can be used to impose a bias on a deep model similar to the spatial bias imposed by convolutions on an image. We explore the usage of Graph Convolutional Neural Networks coupled with dropout and gene embeddings to utilize the graph information. We find this approach provides an advantage for particular tasks in a low data regime but is very dependent on the quality of the graph used. 
We conclude that more work should be done in this direction. We design experiments that show why existing methods fail to capture signal that is present in the data when features are added which clearly isolates the problem that needs to be addressed. 
\end{abstract}

\section{Introduction}

Applications in precision oncology, such as survival analysis and cancer subtype detection, use regression on gene expression data or hand crafted differential expression models. 
Particular genes are usually selected to build prediction models to avoid the impact of spurious correlations between genes expression levels on the predicted labels.
An experienced bioinformatician would know the genes related to a particular tissue or disease given online repositories of gene function and interaction.
In this work we explore how this information can be used to bias the feature construction in a deep learning model.

\begin{figure}[t]
\centering
    \begin{subfigure}[t]{0.5\columnwidth}
    \caption{Median dataset size}
        \includegraphics[width=\columnwidth]{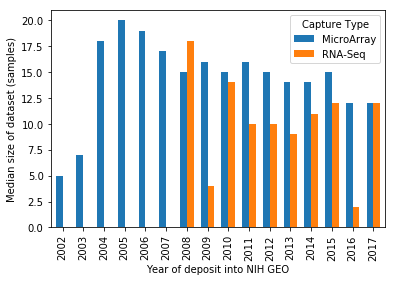}
        
    \end{subfigure}%
    \begin{subfigure}[t]{0.5\columnwidth}
     \caption{Number of datasets added}
        \includegraphics[width=\columnwidth]{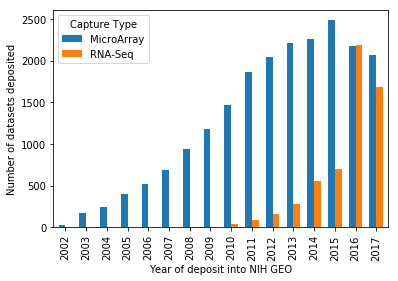}
    \end{subfigure}
    \vspace{-10pt}
    \caption{In the NIH GEO database the median dataset size as well as the number of datasets added each year.
    \vspace{-15pt}}
    \label{fig:geostats}
    
\end{figure}

Due to high cost of data acquisition in biology, most predictive models must be trained in the ``low data" regime. Thousands of studies each year are deposited into the NIH Gene Expression Omnibus (GEO) \footnote{https://www.ncbi.nlm.nih.gov/geo/} database and the number of new samples is rising almost monotonically as seen in Figure  \ref{fig:geostats}. However, the median dataset size remains stagnant. In order for deep learning models to have an impact on biological research we must improve their performance in low data regimes. 

Gene expression data produced from RNA-Seq or MicroArray is often interpreted as a set of independent variables. However, one can take advantage of the experimental data on protein-protein interactions, transcription factors, and gene co-expression to factorize the probability distribution of genes expression levels. These interactions between genes or their products are represented as graphs \cite{Warde-Farley2010}. 

We explore how these graphs can be used to bias a model and subsequently the representation learned using what is known already about the biological system. The bulk of such work has been previously done using linear models \cite{Zhang2017}. Here we try to extend graph bias to modern deep learning models. This approach helps the model to ignore noise which correlates with a target prediction by chance. Also it can reduce the overall number of parameters, analogous to what is done with Convolutional Neural Networks in images.

\begin{figure*}[t]
    \includegraphics[width=\textwidth]{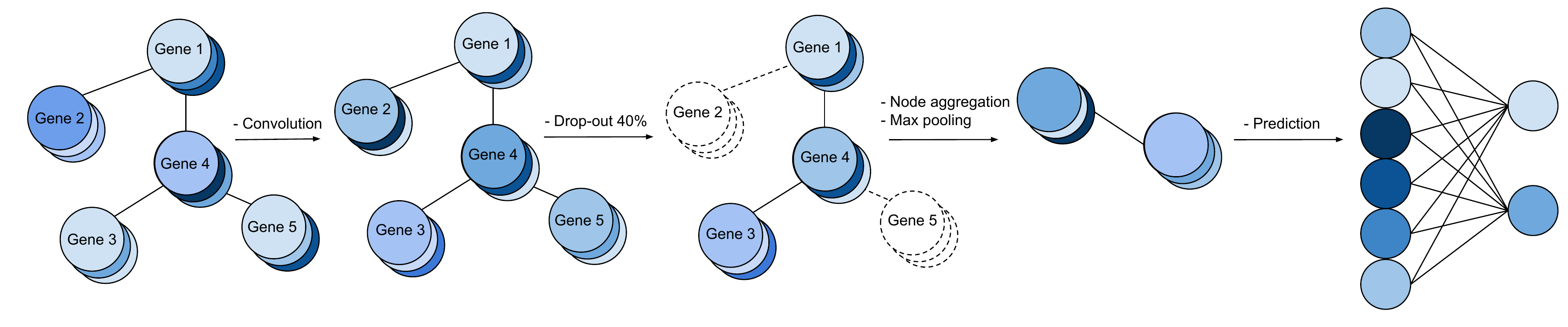}
    \caption{An overview of the Graph Convolutional Network applied to gene expression data. At first each gene is embedded in a graph where neighbors are extracted from prior biological knowledge. After each convolution, the genes are aggregated together based on their connectivity. Finally, a prediction is made from the remaining nodes.}
    \label{fig:pipeline}
\end{figure*}

\section{Gene Graph Convolutions}
\label{GGC}

Most existing work with graph convolutional networks focuses on settings where data is confined to a graph structure, for example points clouds, social networks, or protein structures. With gene expression data, the graphs are complementary to the main task and can be used as a bias. With low numbers of samples, known relationships between variables can help a model avoid learning spurious correlations. 

\textbf{Via regularization:} The method proposed by \citet{Min2016} is to regularize the weights of a Network-regularized Sparse Logistic Regression (NSLR) based on the connectivity of the nodes found in the interaction graph. This is achieved by adding a regularization term using the graph Laplacian $L$ to the logistic regression loss function $\lambda |w|^T L |w|$. This regularization encourages the weights associated with nodes that have a high number of interactions to remain important.

\textbf{Via convolution:} We can also use the structure of the graph as a bias. By performing convolution operations on a node to incorporate information from its neighbors we can extract and propagate the features along the edges of the graph, similar to what happens inside a Convolution Neural Network with adjacent pixels. This convolution over the features $X$ with parameters $\theta$  (denoted $\theta * X$) is not trivial when the structure of the graph is highly complex. \citet{Bruna2013} explored the use of highly sparse Multi-Layer Perceptrons (MLP) where each feature is only linked to its neighbours. They also used a spectral convolution, by projecting the parameters $\theta$ into the spectral space of the Laplacian matrix:
$$
X^{l+1} = \theta * X^l = U diag(\theta) U^T X^l
$$
where $U$ contains the eigenvectors of the Laplacian $L$, $X^l$ and $X^{l+1}$ are the nodes features at layer $l$ and $l+1$ respectively.

However, the full projection of the eigenvectors represent paths of infinite length and will therefore take into account all nodes at once and prevent the network from reasoning about interactions with neighbors. This means that no locality would be present in the convolution, which makes the interpretability and the sharing the parameters more difficult.

To obtain locality in the convolution we can utilize methods in \citet{Defferrard2016} and \citet{Kipf2016} and approximate the convolution to the first neighbouring nodes (paths of length 1). With $A' = A + I_N$ ($A$ being the adjacency matrix) and $D'_i = \sum_j A'_{ij}$, they showed that: %
$$
\theta * X^{l} \approx D'^{-1/2}A'D'^{-1/2}X^l\theta = \tilde{A}X^l\theta
$$%
Where $X^{l} \in \mathbb{R}^{n\times c}$, $\tilde{A} \in \mathbb{R}^{n\times n}$, $\theta \in  \mathbb{R}^{c\times o} $, and $n$ the number of nodes, $c$ the input feature size, and $o$ the output feature size. This leaves us with only $c\times o$ parameters to learn for one specific layer.

This approach does not allow us to have different types of interactions, since all nodes are aggregated before any transformation is done. 
While it is possible to have different sets of parameters for different interactions like in \citet{Bruna2013}, gene interaction graphs do not contain the specific type of interactions between genes. For this reason, we followed \citet{Hamilton2017} and added a skip connection at each convolution layer, which essentially preserves two kinds of signals: the neighborhood and the node itself. 
The full convolution is then followed by an activation function ($ReLU$ in our case) and an aggregation clustering method to reduce the number of nodes at each layer: 
$$
    X^{l+1} = Aggregate(\sigma(\tilde{A}X^l\theta_1 + X^l\theta_2))
$$
In this work hierarchical clustering is used based on the node connectivity in the interaction graph to reduce the number of nodes by half after each convolution. A max pooling is then performed on each resulting cluster. At the last layer, the remaining nodes are concatenated together and fed into a linear layer to make the final prediction. The input $X^0$ of the model are gene embeddings, learned during training, and scaled by their corresponding expression level.

To help with the low amount of data, we also use Drop-out \cite{Srivastava2014}. After every convolutional layer, each node has a 40\% chance of being dropped. This results in the model not being able to rely on a specific node and has to spread the important information across the network, which in turn can make the learning of important features easier. An overview of the model can be seen in Figure \ref{fig:pipeline}.

\section{Experiments}

This work uses 10,459 RNA-Seq samples from the TCGA PANCAN database \cite{CancerGenomeAtlasResearchNetwork2013} spanning multiple tissues and measuring 16,300 genes for each sample. Most samples have been diagnosed with some form of cancer but many healthy examples are also included.

While our ultimate goal is to predict clinical attributes using all gene signals, we have not observed conclusive evidence of deep neural networks outperforming logistic regression baselines in this setting.

To allow us to obtain feedback when comparing different methods, we constructed an experimental setting where we can reduce the difficulty by using only a subset of all genes which are known to be related to the task at hand. Ideally this task would be cancer subtype or phenotypical trait but this does not seem possible for the following reasons: 1. we cannot make any assumptions on the relevant genes 2. we cannot guarantee that any complex relationship is necessary to solve the task.

\begin{figure}[!h]
\floatbox[{\capbeside\thisfloatsetup{capbesideposition={left,top},capbesidewidth=2.5cm}}]{figure}[\FBwidth]
{\caption{This plot illustrates the construction of the single gene inference task by adding nodes based on their distance from the node we want to predict. This graph is for the gene S100A8. The higher the distance the lighter the color.}\hspace{-10pt}\label{fig:shell}}
{\includegraphics[width=5.5cm]{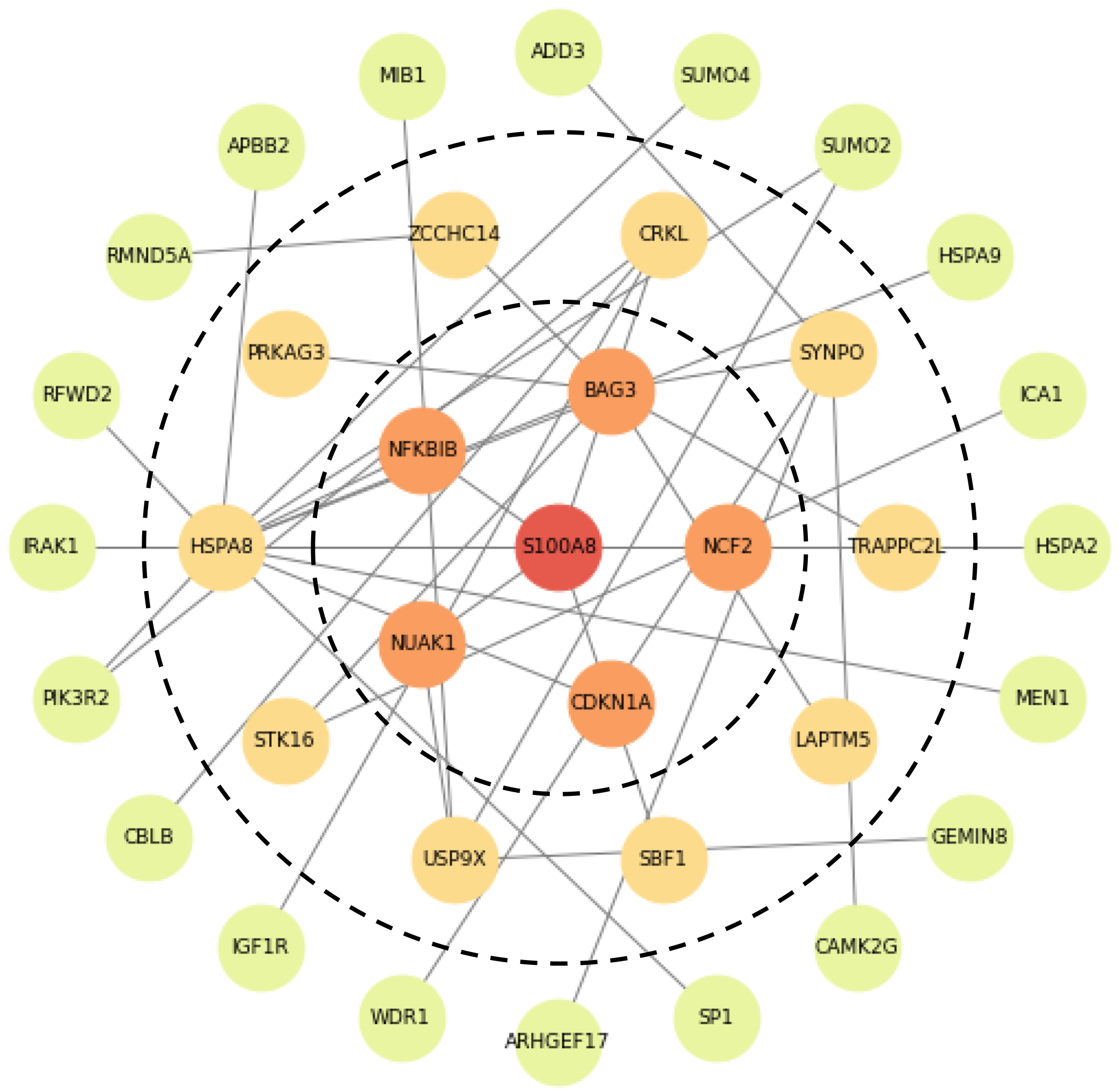}}
\end{figure}

The setting is as follows: we select a specific gene and convert its real-valued expression level to a binary variable representing if it is over or under expressed compared to the mean value for that gene in the 10,459 TCGA samples. This allows a simple binary prediction which we can evaluate using AUC. We then try to predict this value based on the expression level of other genes, not unlike the gene inference task done by \citet{Chen2016g}. To vary the difficulty of the task, we choose different subsets of input genes, based on their connectivity as shown in Figure \ref{fig:shell}. If nodes are of equal distance they are sorted for repeatable experiments. In the easiest setting, we predict the gene over/under expressivity by providing a classification model with the expression values of its  closest neighbors and then add more and more gene expression values until all 16k genes are considered by the model.

\begin{figure}[!h]
\centering
\includegraphics[width=0.9\columnwidth]{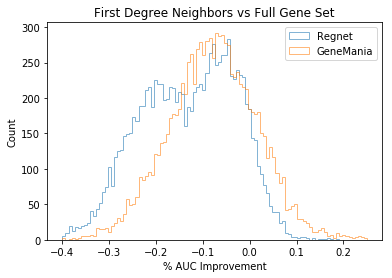}
    \vspace{-10pt}
    \caption{For each graph we train two MLPs to predict each of the 16k genes. One uses all genes and the other uses only the first degree neighbors in the graph. We show the difference in AUC between the models. If a gene has no neighbors then the model predicts 50\%. Genes with a $\%$AUC improvement $> 0$ were better predicted when only considering the first degree neighbors.}
    \label{fig:gapdist}
\end{figure}

\begin{figure*}[!t]
\centering
    \begin{subfigure}[b]{1.0\textwidth}
    \caption{Genes which demonstrate signal exists when noise is absent}
    \label{fig:singlegeneinf:a}
    \begin{subfigure}[t]{0.33\textwidth}
        \includegraphics[width=\textwidth]{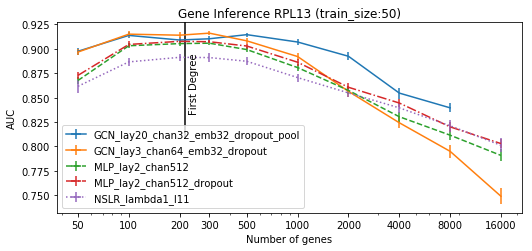}
    \end{subfigure}%
    \begin{subfigure}[t]{0.33\textwidth}
        \includegraphics[width=\textwidth]{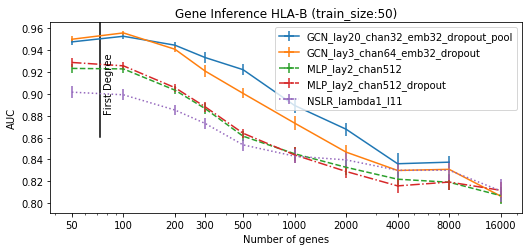}
    \end{subfigure}%
    \begin{subfigure}[t]{0.33\textwidth}
        \includegraphics[width=\textwidth]{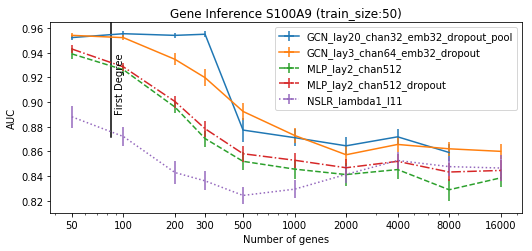}
    \end{subfigure}%
    \vspace{-10pt}
    \begin{subfigure}[t]{0.33\textwidth}
        \includegraphics[width=\textwidth]{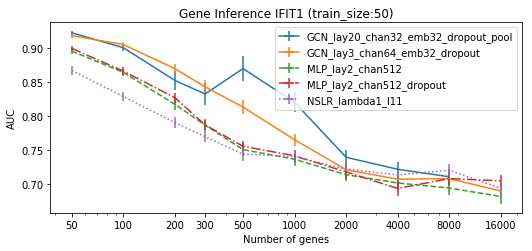}
    \end{subfigure}%
    \begin{subfigure}[t]{0.33\textwidth}
        \includegraphics[width=\textwidth]{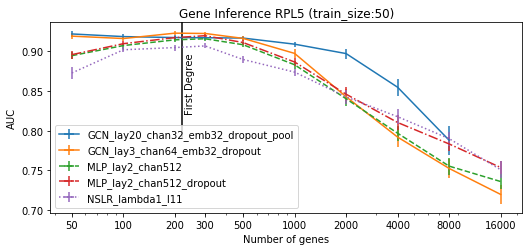}
    \end{subfigure}%
    \begin{subfigure}[t]{0.33\textwidth}
        \includegraphics[width=\textwidth]{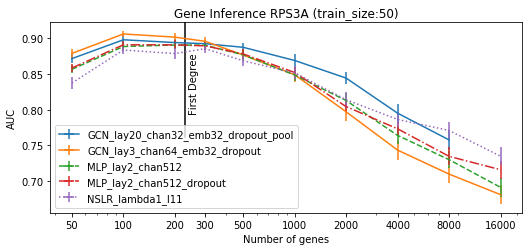}
    \end{subfigure}%
    \end{subfigure}
    \begin{subfigure}[b]{1.0\textwidth}
    \caption{Genes which do not appear impacted}
    \label{fig:singlegeneinf:b}
    \begin{subfigure}[t]{0.33\textwidth}
        \includegraphics[width=\textwidth]{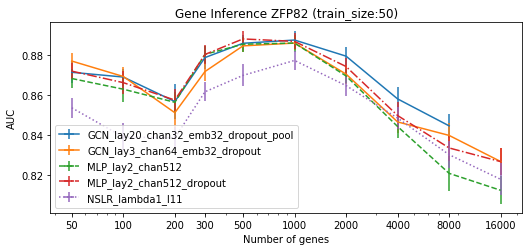}
    \end{subfigure}%
    \begin{subfigure}[t]{0.33\textwidth}
        \includegraphics[width=\textwidth]{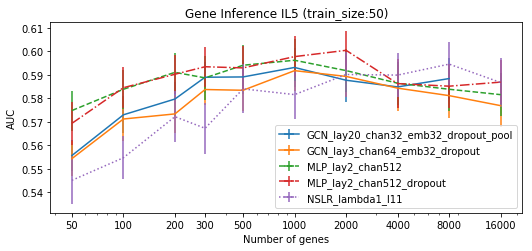}
    \end{subfigure}%
    \begin{subfigure}[t]{0.33\textwidth}
        \includegraphics[width=\textwidth]{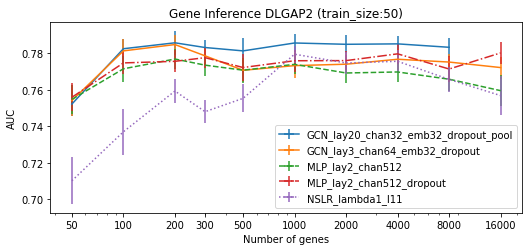}
    \end{subfigure}%
    \end{subfigure}

    \caption{
    We perform experiments to illustrate why predicting labels using all genes (like cancer subtype or other clinical labels we care about) does not work well. We predict if a single gene value is over/under expressed given its neighbors in the GeneMania gene-gene interaction graph. For each plot the target gene varies and the number of gene signals given to each model is increased starting from the small set of neighboring genes and increased to contain all genes in the dataset. 20 trials are used and error bars are shown using standard error. A vertical bar is omitted when the number of first degree neighbors are less than 50.
    }
    \label{fig:singlegeneinf}
\end{figure*}

\subsection{Quality of graphs}
This work explores two public undirected graph datasets containing a mixture of protein-protein interaction and gene co-expression data. The first, GeneMania \cite{Warde-Farley2010}, is a combination of previously published protein-protein interaction and co-expression graphs that contains 264,657 edges covering 16,297 genes. The second, RegNetwork \cite{liu2015regnetwork}, is composed of experimental and predicted up/down regulating effects between genes and includes information from KEGG \cite{Kanehisa2017}. RegNetwork contains 247,848 edges covering 7,220 nodes.

The performance of these graphs is compared in Figure \ref{fig:gapdist}. The percentage of genes where there is an improvement using first degree neighbors is a minority. However, it can yield a $>20\%$ AUC improvement in these cases. Specifically, when using the RegNetwork graph the expression value of $6.25\%$ of target genes were better predicted by their first degree neighbors than the full gene set. This percentage climbs to $13.41\%$ for the GeneMania graph.

Predictions made with the GeneMania graph outperformed those made with RegNetwork even though RegNetwork has almost twice as many edges per node. This finding suggests that simply merging graphs will not yield improved performance. Future work should experiment with directed graphs and multiple edge-types.

\subsection{Robustness to noise}

As discussed earlier, naively comparing models with the whole input is prone to failure. In order to study at what point these models fail we vary the features considered between 50 and all 16k genes. We aim to observe the potential of a model helped by knowledge of a subset of related neighboring genes and how that signal is obscured as we include more and more unrelated genes. In Figure \ref{fig:singlegeneinf} we show the AUC of different models plotted side by side. We compare a MLP (2 layers, 512 channels, ReLU) with and without dropout, a Sparse Logistic Regression model with L1 and network regularization (SLR) and two Graph Convolution Network (GCN), as described in section \ref{GGC}. A simple GCN (3 layers, 64 channels, 32 dim embeddings), and a bigger GCN with 4 pooling layers (24 layers, 32 channels, 32 dim embeddings). The GeneMania graph was used for this task. Each experiment is performed 20 times where each run selects 50 random samples for training,  50 for validation and early stopping, and 1000 for testing/reporting. We were not able to compute the clustering for all 16k genes so pooling cannot be done at that scale.

Three categories of model performance are observed. In Figure \ref{fig:singlegeneinf:a} using neighbors provides an improvement demonstrating that signal exists in the data but these methods are unable to construct features when all genes are considered. This confirms the necessity of doing experiments with a gradual level of difficulty, in order to adequately compare the different models. We observe that Graph Convolutional approaches perform better than a MLP or Logistic Regression model. In Figure \ref{fig:singlegeneinf:b} we observe that models performance is either consistent or increases as neighbors are added, demonstrating a limitation of using graph information, as the most relevant nodes were not in the immediate neighborhood.

\section{Conclusion}

In this paper we explore the difficulty in using deep learning models on gene expression data. We provided an experimental setup where we can evaluate different methods with an increasing level of difficulty. We demonstrate that gene-gene interaction graphs can be utilized in a deep model with Graph Convolutional Networks and that they perform well on the single gene inference task when compared to MLP or logistic regression models.

\section*{Acknowledgements}
This work is partially funded by a grant from the U.S.
National Science Foundation Graduate Research Fellowship
Program (grant number: DGE-1356104) and the Institut
de valorisation des donnees (IVADO). This work utilized the supercomputing facilities managed by the Montreal Institute
for Learning Algorithms, NSERC, Compute Canada,
and Calcul Quebec.

\bibliography{graphs,otherpapers}
\bibliographystyle{icml2018-nopagenum}

\end{document}